\documentstyle[12pt]{article}

\def\theequation{\thesection\arabic{equation}}

\title{\bf A New Approach to the Yukawa Puzzle}
\author{G.C. Branco,\and D. Emmanuel-Costa,\and J.I. Silva-Marcos
\and\bigskip\bigskip\bigskip\and               
{\it Centro de F\'\i sica das Interac\c c\~oes Fundamentais, CFIF,}\and 
{\it Departamento de F\'\i sica, 
Instituto Superior T\'ecnico,}\and{\it Av. Rovisco Pais, 1096
Lisboa Codex, Portugal }}

\date{August 19996}

\newcommand{\noi}{\noindent}
\newcommand{\seccao}[1]{\setcounter{equation}{0} \section{\bf \Large #1}}
\newcommand{\imp}[2]{\bigskip \bigskip {\bf \large \underline{#1} 
#2}\bigskip}
\newcommand{\UM}{1\>\!\!\!\mbox{{\rm I}}}
\newcommand{\diag}{\mbox{\rm diag}}
\newcommand{\ph}{{\mbox{\tiny phase}}}

\newcommand{\dd}[1]{#1\rightarrow \ \delta}
\newcommand{\cc}[1]{#1\rightarrow \ \chi}  
\newcommand{\va}[2]{|V_{#1 #2}|=}
\newcommand{\raiz}[2]{\sqrt{\frac{#1}{#2}}}

\begin{document}

\maketitle
\begin{abstract}

We do a systematic analysis of the question of calculability of CKM 
matrix elements in terms of quark mass ratios, within the framework of 
the hypothesis of universality of strength for Yukawa couplings (USY), 
where all Yukawa couplings have equal moduli, and the flavor dependence 
is only in their phases. We 
use the fact that the limit $m_u=m_d=0$ is specially simple in 
USY, to construct the various ans\"atze. It is shown that the 
experimentally observed CKM matrix can be obtained within USY ans\"atze 
corresponding to simple relations among phases of Yukawa couplings. 
Within USY, one finds 
a natural explanation why Cabibbo mixing is significantly larger than the 
other CKM mixings. In the most successful of the USY ans\"atze, one 
obtains in leading order:
$|V_{us}|\,=\,\sqrt{m_d/{m_s}}$ ; $ |V_{cb}\,|=\,\sqrt{2}(m_s/m_b)$
$ |V_{ub}|\,=\, (1/\sqrt{2})\sqrt{m_dm_s/m_b^2}$
; $|V_{td}|\,=\, 3\,\,|V_{ub}|$.
We study the behavior of this USY ansatz under the renormalization 
group.

\end{abstract}
\pagebreak

\seccao{Introduction}

In the standard model (SM), the flavor structure of Yukawa interactions is
not constrained by any symmetry, thus leading to arbitrary Yukawa
couplings consisting of three 3x3 complex matrices. This arbitrariness is
used to fit the lepton and quark masses, as well as the four physical
parameters of the Cabibbo-Kobayashi-Maskawa (CKM) matrix. 

Finding a deeper insight into the pattern of fermion masses and mixings,
is one of the major outstanding problems in particle physics. In the past,
there have been various attempts at relating the pattern of CKM mixings to
the quark mass ratios \cite{r1}. In most schemes, one assumes that some of
the Yukawa matrix elements vanish, which leads to testable relations
between quark masses and the elements of the CKM matrix. Recently, a
classification has been done \cite{r2} of all Hermitian matrices with
"textures zeros", which conform to our present knowledge on quark masses
and CKM matrix elements. 

In this paper, we will analyse in a systematic way the question of
calculability of the CKM matrix elements within the framework \cite{r3} of
universality of strength for Yukawa couplings (USY). 
In USY all the Yukawa couplings of the quarks have  
equal moduli, but differ in their phases.
A physical motivation
for the USY hypothesis may be found in the following observations: 

(i) The Yukawa interactions are the only couplings of the SM which can be 
complex. All other couplings are constrained to be real by hermiticity.

(ii) Most of the arbitrary parameters of the SM arise precisely from the 
Yukawa couplings.

If one assumes that the above two features are somehow related, one is
naturally led to the USY hypothesis, i.e. to the idea that the
arbitrariness of Yukawa couplings results from the fact that they can be
complex, with flavor-dependent phases, but universal strength.

In our study of the problem of calculability of the CKM matrix elements 
within the USY framework we will start by considering the limit where the 
first generation of quarks is massless. It has been previously pointed 
out \cite{r4} that within USY all solutions leading to $m_u=m_d=0$, can be 
classified. In this paper, we construct various new ans\"atze based on USY by 
considering the different ways of departing from this limit which lead to 
calculability of CKM matrix elements in terms of quark mass ratios. The 
ans\"{a}tze correspond to simple relations among the USY phases. In the 
search for these relations among phases, we will use as guiding 
principles, on the one hand simplicity and on the other hand some of the 
main features of the experimentally observed pattern of CKM mixings, 
namely the fact that $|V_{us}|$ is of order $(m_d  /m_s)^{1/2}$, while 
$|V_{cb}|$ is of order $(m_s/m_b)$. We offer a generic argument how these 
relations can be obtained and show that they naturally arise within the 
USY framework.

The paper is organized as follows: in the next section, we will set our
notation and characterize the parameter space of the Yukawa couplings
within USY. In section 3, we will address in a systematic way the question
of calculability of CKM matrix within the USY framework. In section 4 we
confront the various ans\"{a}tze with the experimental value of quark
masses and mixings. A study of the behavior under the renormalization
group of a particularly successful USY ansatz is presented in section 5.
In section 6, we compare the analysis of the present paper with the
texture zeros approach \cite{r2} to the Yukawa puzzle. Our conclusions are
presented in section 7. In the appendix, we show some special features of
USY. In particular, we point out that within USY, a successful prediction
for $|V_{us}|$ requires three generations. Indeed we show that for two
generations, USY leads either to an arbitrary non-calculable Cabibbo angle
or to an unrealistic relation
$|\theta_C|=\frac{m_d}{m_s}\pm\frac{m_u}{m_c}$. \bigskip \seccao{USY
parameter space}

For completeness and in order to settle our notation, we describe 
the parameter space of Yukawa couplings in the USY framework. We  
assume that there are two Higgs doublets $\Phi_u$, $\Phi_d$ which give 
mass to the up and down quarks respectively, through Yukawa couplings of 
universal strength. All the flavor dependence is contained in the phases 
of Yukawa couplings and therefore the up and down quark masses have the 
form:

\begin{equation}
[M_u ]_{ij}=c_u \, \exp[i\phi^u_{ij}] 
\quad \quad \quad \quad 
[M_d ]_{ij}=c_d \, \exp[i\phi^d_{ij}] 
\label{eq21}
\end{equation}

\noi with $c_u=\lambda v_u$, $c_d=\lambda  v_d$, where $\lambda$ 
denotes the universal strength of Yukawa couplings and $v_u=<\Phi_u>$,
$v_d=<\Phi_d>$. One can eliminate some of the phases appearing in 
Eq.(\ref{eq21}) by making weak-basis transformations of the type:

\begin{equation}
\begin{array}{r}
M_u \rightarrow M'_u=K^{\dagger}_L\cdot M_u \cdot K^u_R
\\ \\
M_d \rightarrow M'_d=K^{\dagger}_L\cdot M_d  \cdot K^d_R 
\end{array}
\label{eq22}
\end{equation} 

\noi where $K_L$, $K^{u,d}_R$ are diagonal unitary 
matrices, i.e. of the form
$\diag(e^{i\phi_1},e^{i\phi_2},e^{i\phi_3})$. 
Obviously the transformations of Eq.(\ref{eq22}) keep the USY 
form of Yukawa couplings. 
In general it is not possible to obtain
$M'_u$, $M'_d$ Hermitian, while maintaining the USY form. Actually
it can be readily verified that the Hermitian USY matrices lead to 
unrealistic mass matrices.
However, it can be easily seen that, by making the 
weak-basis transformations of Eq.(\ref{eq22}), one can transform any USY 
matrices $M_u  $, $M_d  $ given by Eq.(\ref{eq21}), into the form:

\begin{equation}
M_u =c_u \, \left( 
\begin{array}{ccc}
e^{ip_u} & e^{ir_u} & 1 \\
e^{iq_u} & 1 & e^{it_u} \\
1 & 1 & 1
\end{array}
\right) \quad M_d =c_d \, K^\dagger\cdot  \left( 
\begin{array}{ccc}
e^{ip_d} & e^{ir_d} & 1 \\
e^{iq_d} & 1 & e^{it_d} \\
1 & 1 & 1
\end{array}
\right)\cdot K
\label{eq23}
\end{equation}

\noi where $K=\diag(1,e^{i\alpha_1},e^{i\alpha_2})$.
It is clear from Eq.(\ref{eq23}) that the phases $\alpha_i$ will affect
the CKM matrix but not the quark mass spectrum which depends only on
$p_{u,d}$; $q_{u,d}$;  $r_{u,d}$; $t_{u,d}$. It is useful to introduce 
dimensionless Hermitian matrices defined by 

\begin{equation}  
H_u=\frac1{3c^2_u}\,\, M_u   M_u^\dagger
\quad \quad \quad \quad   
H_d=\frac1{3c^2_d}\,\, M_d   M_d^\dagger
\label{eq23a}
\end{equation}

\noi and the related pure phase mass matrices

\begin{equation}
M^{\ph}_u=\,\,\left(
\begin{array}{lll}
e^{ip_u} & e^{ir_u} & 1 \\
e^{iq_u} & 1 & e^{it_u} \\
1 & 1 & 1
\end{array}
\right) \quad M^{\ph}_d=\,\,\left(
\begin{array}{lll}
e^{ip_d} & e^{ir_d} & 1 \\
e^{iq_d} & 1 & e^{it_d} \\
1 & 1 & 1
\end{array}
\right)
\label{eq23b}
\end{equation}

\noi In these pure phase matrices, the constants $c_{u,d}$ and the
diagonal unitary matrix $K$ are not included in order 
to simplify the presentation.
The eigenvalues $\lambda^{u,d}_i$ of $H_{u,d}$ are also dimensionless, 
and related to the squared quark masses $m^2_i$ by
$\lambda_i=3m^2_i/(m^2_1+m^2_2+m^3_3)$. The coefficients of the
characteristic equations are given by the trace $tr(H)$, the second
invariant $\chi(H)$, and the determinant $\delta(H)$ of the matrices
$H$

\begin{equation}
\begin{array}{lll}
tr\: H & \equiv & \lambda_1 + \lambda_2 + \lambda_3=3\\ \\ \\ 
\chi(H) & \equiv & 
\lambda_1\lambda_2+\lambda_1\lambda_3+\lambda_2\lambda_3%
=\frac49 [
\sin^2(\frac p2)+\sin^2(\frac q2)+\sin^2(\frac r2)+\sin^2(\frac 
t2) + \\ \\
& & \sin^2(\frac {r+t}2)+\sin^2(\frac {p-r}2)
+\sin^2(\frac{q-t}2)+ \sin^2(\frac {p-q-r}2)
+\sin^2(\frac {p-q+t}2) ] \\ \\ \\
\delta(H) & \equiv & \lambda_1 \lambda_2 \lambda_3=\frac1{27}|%
det(M^\ph)|^2  \\ \\
& &=\frac1{27}
|e^{ip}+e^{iq}+e^{i(r+t)}-e^{i(p+t)}-e^{i(q+r)}-1|^2=\frac1{27}|A+B|^2
\end{array}
\label{eq25}
\end{equation}

\bigskip

\noi where    

\begin{equation}
A=-(1-e^{ip})(1-e^{it})
\quad \quad \quad \quad
B=e^{it}(e^{i(q-t)}-1)(1-e^{ir})
\end{equation}

The quark mass hierarchy leads to the constraint $\chi(H)\ll 1$ and since 
$\chi(H)$ is the sum of positive definite quantities, the modulus of each 
one of the phases $p$, $q$, $r$ and $t$ has to be small, at most of order 
$\frac{m_2}{m_3}$.

\bigskip
\seccao{Calculability of V$_{CKM}$}

It has been shown \cite{r3}, \cite{r5} that within the USY hypothesis, one
can correctly fit the observed pattern of CKM matrix elements, as well as
the value of quark masses. However, without any further assumptions, the
USY hypothesis has the disadvantage of containing too many free
parameters. In this section, we will make a systematic study of ans\"atze
based on USY, leading to calculability of the CKM matrix elements, in
terms of quark masses. In order to achieve this, each one of the quark
mass matrices $M_u$, $M_d$ should depend only on the over-all constants
$c_u$, $c_d$ and two phases. One will then have a total of six parameters
in $M_u$, $M_d$ which will be fixed by the value of the six quark masses.
As a result the CKM matrix will be a function of quark mass ratios with no
free parameters. 

In our search for these ans\"atze, we will start by considering the limit 
\mbox{$m_u=m_d=0$}. In the USY framework, this limit is specially interesting 
since it is possible to find exactly all solutions [4] leading to 
$det(M_{u,d})=0$. We divide these solutions into two classes:

\begin{equation}
\mbox{\rm Class-I}\left\{\begin{array}{lll}
a) & p=0, & t=q \\
b) & t=0, & r=0 \\
c) & r=p, & q=0 \end{array} \right.
\quad \quad \quad
\mbox{\rm Class-II}\left\{\begin{array}{lll} 
a) & p=0, & r=0 \\
b) & q=0, & t=0 \\
c) & r=p-q, & t=-r \end{array} \right.
\label{eq31}
\end{equation}

\noi where in each solution, the omitted parameters are arbitrary, only
constrained by $\chi(H)$, given in Eq.(\ref{eq25}). 

The solutions of Eqs.(\ref{eq31}) can be readily understood from
Eqs.(\ref{eq25}), where the determinant of the pure phase matrix,
$det(M^\ph)$, is written as the sum of two complex numbers A, B. Solutions
Ic) and IIc) correspond to having $A=-B$, while the other solutions
correspond to $A=B=0$. In order to see the distinct physical implications
of these two classes of solutions, it is useful to consider the limit
\mbox{$K=\UM$} (i.e. $\alpha_1=\alpha_2=0$), where $K$ has been defined in
Eq.(\ref{eq23}). We recall that the two phases $\alpha_1$, $\alpha_2$ do
not affect the quark mass spectrum, only entering in the CKM matrix. For
\mbox{$K=\UM$}, and in the limit $m_u=m_d=0$, solutions of Class-II cannot
generate a realistic CKM matrix, while solutions of Class-I can generate a
realistic CKM matrix even in the limit of massless $m_u $, $m_d $. 

In the search for a viable ansatz based on USY, we will consider that
the masses for the first generation are generated through a small
deviation of the limit $det(M_u )=det(M_d  )=0$. We will do all 
calculations exactly, without using perturbation theory. The fact that in USY
this limit is characterized by two conditions on the phases, suggests that
the generation of mass for the first generation can be obtained 
through the relaxation
of one of these conditions. Following this suggestion, one finds the
following cases\footnote{We have not included the case $\{r=p-q\}$, since 
it leads to unrealistic predictions for $V_{CKM}$.}: 

\begin{equation} 
1)\: \{p=0\}; \quad 2)\:\{ t=0\}; \quad 3)\: \{t=q\}; 
\quad 4)\:\{ t=-r\} 
\end{equation}

One can find another set of physically equivalent cases by making the 
interchange $(p,r)\leftrightarrow(q,t)$. For definiteness, we will 
consider next the case $\{p=0\}$. From Eq.(\ref{eq25}), one obtains
for the determinant of the pure phase matrix $M^{\ph}$:

\begin{equation}
det(M^{\ph})=e^{it}(e^{i(q-t)}-1)(1-e^{ir})
\end{equation}

\noi It follows then that:

\begin{equation}
|det(M^{\ph})|\,\,=4\,\,\left|\sin(\frac{q-t}2)\: \sin(\frac r2)\right|
\label{eq34}
\end{equation}

At this stage, taking \mbox{$K=\UM$}, 
we have in each of the original full mass matrices 
$M_u  $, $M_d  $ four
parameters, namely $(c_u,q_u,r_u,t_u)$ and $(c_d,q_d,r_d,t_d)$. In order
to achieve full calculability of the CKM matrix (i.e. having $V_{CKM}$
entirely expressed in terms of quark masses ratios, with no free
parameters), each one of the matrices $M_u  $, $M_d  $ should contain only
three parameters. Therefore we need an extra relation, among the phases 
$q$, $r$, $t$. In looking for such a relation, "simplicity" 
will be our guiding principle.
In particular we will search for relations among $q$, $r$ and $t$ such 
that $det(M^\ph)$ depends only on one phase-parameter. Later,
we will present a heuristic argument in favor of this scenario.
Following this suggestion and taking into account Eq.(\ref{eq34}), we set

\begin{equation}
|q-t|\,\,=\,\,|r|
\end{equation}

\noi thus obtaining:

\begin{equation}
|det(M^{\ph})|\,\,=\,\,4\sin^2(\frac r2)
\label{eq36}  
\end{equation}

\noi or equivalently

\begin{equation}
\sin^2(\frac r2)={\frac 3{4}}\sqrt{3\delta}
\label{eq36a}
\end{equation}

\noi where we used the relation $|det(M^\ph)|\,=3\sqrt{3\delta}$ from
Eq.(\ref{eq25}). If we now insert the relations $p=0, \: (q-t)=r$ in the
expression for $\chi$ given by Eq.(\ref{eq25}), we get: 

\begin{equation}
\chi(H) = \frac49 \left[ \sin^2(\frac {q-r}2)+\sin^2(\frac {q+r}2)+
2\sin^2(\frac q2)+4\sin^2(\frac r2) \right]
\label{eq37}
\end{equation}

Using the identity 
$\sin^2(\frac{q-r}2)+\sin^2(\frac{q+r}2)=2\sin^2(\frac q2)+%
2\sin^2(\frac r2)-4\sin^2(\frac q2)\sin^2(\frac r2)$ we can write:

\begin{equation}
\chi(H) = \frac89\left[ 2\sin^2(\frac q2)+3\sin^2(\frac r2)-
2\sin^2(\frac q2)\sin^2(\frac r2) \right]
\label{eq38}
\end{equation}

Using Eqs.(\ref{eq36a}) and (\ref{eq38}), one finally obtains:

\begin{equation}
\sin^2(\frac q2)=\frac{\frac 9{16}\chi - \frac 98\sqrt{3\delta}}%
{1-\frac 34\sqrt{3\delta}}
\label{eq39}
\end{equation} 

It is worth summarizing what we have accomplished so far. By studying the 
limit $det(M_{u,d}   )=0$, we motivated an ansatz where $p=0$, 
$t=q-r$, which led to the following results:

 (i) Each one of the mass matrices $M_u  $, $M_d  $ depends on three
parameters $\{ c_{u,d}, \: \: q_{u,d}, \: \: r_{u,d} \}$. 

(ii) The parameters $c_{u,d}$ are overall constants which are fixed by 
the sum of quark squared masses through the relation 
$c^2=\frac19(m_1^2+m_2^2+m_3^2)$, while $|r|$ and $|q|$ are fixed by 
Eqs.(\ref{eq36a}) and (\ref{eq39}), respectively.

So far, we have been only concerned with the quark mass spectrum. Next, we
turn to the CKM matrix and present the previously mentioned heuristic
argument justifying why the situation described above is potentially
useful to
obtain a calculable and physically interesting CKM matrix. The argument
goes as follows: let us consider an ansatz, not necessarily based on USY,
leading to mass matrices which depend only on two parameters $\rho_1$,
$\rho_2$, such that, at least to leading order, $|\rho_1|^2\propto
|det(M^{\circ})|$ while $|\rho_2|^2\propto\chi(H^\circ )$, where $M^\circ$
and $H^\circ=M^\circ M^{\circ \dagger}$ are the dimensionless matrices 
defined by $M^\circ=M/\sqrt{tr(H)}$. Then, if the mixing $|V_{12}|$ between
the first and second generations is proportional to the ratio
$|\rho_1/\rho_2|$ one would have: 

\begin{equation}
|V_{12}|\propto|\frac{\rho_1}{\rho_2}|\propto
\sqrt{\frac{|det(M^\circ)|}{\chi(H^\circ)}}
\label{eq310}
\end{equation}

Taking into account that the definition of $M^\circ$ implies that
$|det(M^\circ)|\propto\frac{m_1m_2m_3}%
{(m_1^2+m_2^2+m_3^2)^{3/2}}$ while 
$\chi(H^\circ)\propto\frac{m_1^2 m_2^2+m_2^2 m_3^2+m_1^2 m_3^2}%
{(m_1^2+m_2^2+m_3^2)^2}$ one obtains in leading order:

\begin{equation}
|V_{12}|\propto\sqrt{\frac{m_1}{m_2}}
\label{eq311}
\end{equation}

The above qualitative argument will be used as a guideline to do a
systematic search for ans\"{a}tze within the USY hypothesis which have the
feature of leading to $|V_{12}|\propto\sqrt{\frac{m_1}{m_2}}$ while
$|V_{23}|\propto\frac{m_2}{m_3}$. Next, we analyse in some detail two
examples of ans\"atze constructed following the guidelines described
above. 

\imp{Ansatz I:}{$\{p=0$, $t=q-r\}$}

This is the ansatz we have just constructed, leading to mass matrices 
of the form:

\begin{equation}
M_{u,d}=c_{u,d}\left(
\begin{array}{lll}
1 & e^{ir} & 1 \\
e^{iq} & 1 & e^{i(q-r)} \\
1 & 1 & 1
\end{array}
\right)
\label{eq312}
\end{equation}

In order to diagonalize the Hermitian matrices $H_u$, $H_d$ of
Eq.(\ref{eq23a}) and to obtain the CKM matrix, it is useful to make first
a change of weak basis: 

\begin{equation}
H_u \rightarrow H'_u=F^\dagger\cdot H_u \cdot F \quad ; \quad 
H_d \rightarrow H'_d=F^\dagger\cdot H_d \cdot F
\label{eq313}
\end{equation}

\noi where $F$ is given by:

\begin{equation}
F=\left(
\begin{array}{ccc}
\frac1{\sqrt2} & \frac{-1}{\sqrt6} & \frac1{\sqrt3} \\
0 & \frac2{\sqrt6} & \frac1{\sqrt3} \\
\frac{-1}{\sqrt2} & \frac{-1}{\sqrt6} & \frac1{\sqrt3} 
\end{array}
\right)
\label{eq314}
\end{equation}

\noi Through this weak-basis transformation we change from a "democratic"
to a "heavy" basis. The matrices $H'_u$, $H'_d$ are diagonalized by
unitary transformations: 

\begin{equation}
U_u^\dagger \cdot H'_u \cdot U_u=D_u \quad ; \quad
U_d^\dagger \cdot H'_d \cdot U_d=D_d
\label{eq315}
\end{equation}  

\noi where $D=\diag(\lambda_1,\lambda_2,\lambda_3)$. The CKM matrix is
then given by: \mbox{$V_{CKM}=U_u^\dagger U_d$.} Since for this ansatz
$\sin^2(q/2)$ and $\sin^2(r/2)$ can be expressed in terms of quark mass
ratios as in Eqs.(\ref{eq36a}) and (\ref{eq39}), an exact analytical
solution to the eigenvalue equation: 

\begin{equation}
(H'-\lambda_i \UM)\:\vec{v_i}=0
\label{eq316}
\end{equation}

\noi can be found. One obtains in leading order:

\begin{equation}
\begin{array}{llll}
|U^d_{12}|=\sqrt{\frac{m_d}{m_s}} & & & |U^d_{23}|=\sqrt2 \:\frac{m_s}{m_b} 
\\ \\
|U^d_{13}|=|U^d_{12}||U^d_{23}|/2=\frac1{\sqrt2}\sqrt{\frac{m_d m_s}{m_b^2}}
& & &
|U^d_{31}|=3|U^d_{13}|
\end{array}
\label{eq317}
\end{equation}

Of course, entirely analogous expressions hold for $U^u_{ij}$. The values of
$U^u_{ij}$, $U^d_{ij}$ given by Eq.(\ref{eq317}) can be obtained in a much 
simpler way, by noting that from Eq.(\ref{eq315}) one gets:

\begin{equation}
\begin{array}{ll}
H'_{23} & =U_{21}U^\ast_{31}\lambda_1 + U_{22}U^\ast_{32}\lambda_2 
+ U_{23}U^\ast_{33}\lambda_3  \\
& \approx U_{23}U^\ast_{33}\lambda_3
\end{array} 
\label{eq318}  
\end{equation}

On the other hand, $H'_{23}$ is readily evaluated from 
Eqs.(\ref{eq312}) and (\ref{eq313}), and one obtains in leading order:

\begin{equation}
|H'_{23}|=\frac{2\sqrt2}3\: |q|
\label{eq319}
\end{equation}

Taking into account that $\lambda_3=3m_3^2/(m_1^2+m_2^2+m_3^2)%
 \approx 3$ and
$|U_{33}|\approx 1$ it follows from Eqs.(\ref{eq318}) and (\ref{eq319}) that 
in leading order: 

\begin{equation}
|U_{23}|\,\,=\frac13\:|H'_{23}|\:=\:\frac{2\sqrt2}9\: |q| \:=\: 
\sqrt2\:\frac{m_2}{m_3} 
\end{equation}

\noi where we used $|q|\approx (9/2)(m_2/m_3)$ from Eq.(\ref{eq39}). Similarly, 
from $H'_{13}$, written as:

\begin{equation}
\begin{array}{ll}
H'_{13} & =U_{11}U^\ast_{31}\lambda_1 + U_{12}U^\ast_{32}\lambda_2
+ U_{13}U^\ast_{33}\lambda_3  \\
& \approx U_{13}U^\ast_{33}\lambda_3
\end{array}
\end{equation}

\noi and taking into account from Eqs.(\ref{eq312}) and (\ref{eq313}) that

\begin{equation}
|H'_{13}|\,\,=\,\,\frac1{\sqrt6} \: |r|
\end{equation}  

\noi it follows that:

\begin{equation}
|U_{13}|\:=\:\frac13\,\,|H'_{13}|=\frac1{3\sqrt6}\:|r|=
\frac1{\sqrt2}\sqrt{\frac{m_1m_2}{m_3^2}}
\end{equation}

\noi where we used the fact that Eq.(\ref{eq36a}) implies 
$r\approx 3\sqrt3\sqrt{m_1 m_2/m_3^2}$. 

In an analogous way one can derive the leading order value of $|U_{12}|$ and
then using unitarity obtain $|U_{31}|$. These values agree with
those presented in Eq.(\ref{eq317}). 

A qualitative understanding of these predictions for $V_{CKM}$ can be 
obtained by viewing this ansatz as a small perturbation of the zero mass
solution Class-Ia) of the equation $det(M)=0$ corresponding to mass 
matrices of the form:

\begin{equation}
M=\,\, c\,\,\left(
\begin{array}{lll}
1 & e^{ir} & 1 \\
e^{iq} & 1 & e^{iq} \\
1 & 1 & 1
\end{array}
\right)
\label{eq324}
\end{equation}

Since in this limit $m_u=m_d=0$, it is straightforward to find the mass 
eigenstates. After making the weak-basis transformation of Eq.(\ref{eq313}) 
one obtains for small $|r_d/q_d|$, in leading order:

\begin{equation}
\begin{array}{llll}
|U^d_{12}|=\frac{\sqrt3}2\:|\frac{r_d}{q_d}| & & &
|U^d_{23}|=\sqrt2\:\frac{m_s}{m_b}
 \\ \\
|U^d_{13}|=|U^d_{12}||U^d_{23}|/2 & & &
|U^d_{31}|=3\:|U^d_{13}|
\end{array}
\label{eq325}
\end{equation}

This shows that already in the limit $m_u=m_d=0$, the main features of the
Ansatz-I are manifest: there is a clear distinction between the mixing of
the second and third generations and the mixing of the first and second
generations. While $|U_{23}|$ is proportional to a small parameter $|q|$,
which then leads to the prediction $|U_{23}|=\sqrt2(m_s/m_b)$, $|U_{12}|$
is proportional to the ratio of two small parameters $|r|$ and $|q|$.
Thus, one finds a natural explanation why $|U_{12}|$ is much larger than
the other mixings. The Ansatz-I of Eq.(\ref{eq312}) can thus be viewed as
a small perturbation of the mass matrix of Eq.(\ref{eq324}) whose main
effect, is generating mass for the first family and fixing the ratio
$|r/q|$ to the value $|r/q|\cong\frac2{\sqrt3}\sqrt{\frac{m_d}{m_s}}$,
thus leading to the successful prediction
$|U_{12}|=\sqrt{\frac{m_d}{m_s}}$. 

We find remarkable the occurrence of the numerical factor $2/\sqrt3$ in
$|r/q|$, which just cancels with the factor $\sqrt3/2$ in
Eq.(\ref{eq325}). 

\imp{Ansatz II:}{$\{t=0$, $r=-q\}$}

So far we have constructed only one ansatz, following the general
procedure described in the beginning of this section. We will construct a
second ansatz, by putting $t=0$ in Eq.(\ref{eq23}) and noting that in this
case one has: 

\begin{equation}
det(M^\ph)=-(1-e^{ir})(1-e^{iq})
\end{equation}

Following our heuristic argument that $|det(M)|$ should depend only of 
one parameter, we put $|r|=|q|$, in particular $r=-q$, to obtain:

\begin{equation}
\begin{array}{c}
|det(M^{\ph})|\,\,=\,\,4\sin^2(\frac r2)
\\ \\
\chi(H)=\frac89[2\sin^2(\frac p2) + 3\sin^2(\frac r2)-2\sin^2(\frac p2) 
\sin^2(\frac r2)] 
\end{array}
\label{eq326}
\end{equation}

\noi and thus 

\begin{equation}
\sin^2(\frac r2)={\frac 3{4}}\sqrt{3\delta} \qquad , \qquad 
\sin^2(\frac p2)=\frac{\frac 9{16}\chi - \frac 98\sqrt{3\delta}}%
{1-\frac 34\sqrt{3\delta}}
\label{eq326a}
\end{equation}

 As in the previous ansatz, $M_{u,d}$ depend each only on three parameters
$c_{u,d}$, $p_{u,d}$ and $r_{u,d}$, which are fixed in terms of quark mass
ratios through Eq.(\ref{eq326a}). An exact solution for the quark
eigenstates can be readily found and one obtains in leading order: 

\begin{equation}
\begin{array}{llll}
|U_{12}|=\sqrt{\frac{m_1}{m_2}} & & &
|U_{23}|=\frac1{\sqrt2}\left[ 1-\sqrt3\sqrt{\frac{m_1}{m_2}}
\right] \frac{m_2}{m_3} \\ \\
|U_{31}|= \sqrt{\frac32}\:\frac{m_1}{m_3} & & &
|U_{13}|=\frac1{\sqrt2}\sqrt{\frac{m_1m_2}{m_3^2}}
\end{array}
\end{equation}

By now, it should be clear the procedure to construct ans\"{a}tze based on
USY, whose main features are predicting $U_{12}\propto\sqrt{m_1/m_2}$ while
$|U_{23}|\propto m_2/m_3$. In Table 1, we present the various ans\"{a}tze,
together with their predictions for $V_{CKM}$, in leading order. 

\seccao{Confronting with experiment} 
 
In the previous section, we have done a systematic search for ans\"{a}tze 
based on USY which can lead to calculability of the CKM matrix. Our 
starting point was looking for ans\"atze which correctly predict the 
values of $|V_{12}|$ and $|V_{23}|$. The emphasis on these two matrix 
elements is justified on a number of grounds. On the one hand, these are the 
two best measured off-diagonal elements of the CKM matrix. On the other 
hand, for three generations, unitarity of the CKM matrix leads to the 
following exact relations:

\begin{equation}
\begin{array}{c}
|V_{21}|^2=|V_{12}|^2-\epsilon\\ \\
|V_{32}|^2=|V_{23}|^2-\epsilon
\end{array}
\end{equation}

\noi where $\epsilon=|V_{31}|^2-|V_{13}|^2$. Due to the smallness of
$\epsilon$ compared to $|V_{12}|^2$ and $|V_{23}|^2$, one has to a good
approximation $|V_{21}|\cong|V_{12}|$ and $|V_{32}|\cong|V_{23}|$.
Therefore, a good prediction for $|V_{12}|^2$, $|V_{23}|^2$ implies also a
good prediction for $|V_{21}|^2$, $|V_{32}|^2$. 

Before making a detailed comparison of the predictions of the various
ans\"{a}tze with the experimental results, the following point is in
order. The predictions for $|V_{ij}|$ presented in Table 1, were obtained
with the assumption $\alpha_1=\alpha_2=0$. We recall that these are the
phases which enter into the most general parametrization of USY in
Eq.(\ref{eq23}) and which do not affect the quark mass spectrum but enter
in the CKM matrix. In the limit of vanishing $\alpha_i$, the various USY
ans\"{a}tze presented in Table 1 lead to full calculability of $V_{CKM}$,
in terms of quark mass ratios with {\em no free parameters}. In the limit
$\alpha_i=0$, only Ansatz-I can correctly predict all elements of
$V_{CKM}$. However, it is clear that the requirement of full calculability
(i.e. no free parameters) is not necessary. Actually, Ansatz-I is rather
unique, since it is the only ansatz which correctly predicts $V_{CKM}$,
without free parameters. Indeed, to our knowledge, none of the numerous
ans\"{a}tze proposed in the literature predict $V_{CKM}$, without free
parameters.  For example, the well known Fritzsch ansatz \cite{r6}
predicts $V_{CKM}$ in terms of quark mass ratios and two arbitrary phases.
These two phases are entirely analogous to the $\alpha_i$ phases which
appear in the USY framework. Analogous arbitrary phases also appear in all
the texture zero ans\"{a}tze which have been recently classified
\cite{r2}. 

Next, we give two examples where the values of $|V_{CKM}|$ are correctly 
predicted. The first example corresponds to Ansatz-I, where 
the CKM matrix is correctly predicted in terms of quark mass 
ratios, without free parameters. The second example corresponds to 
Ansatz-II, where $V_{CKM}$ is predicted in terms of quark mass ratios 
and one free parameter (the phase $\alpha_2$ is set to $\alpha_2=0.0093$).

\bigskip
Input: $m_t^{physical}= 174\, GeV$ and
$$
\left[\begin{array}{lll}
m_u(1\, GeV)=1.0\, MeV   \\ 
m_c(1\, GeV)=1.35\, GeV \\
m_d(1\, GeV)=6.5\, MeV \\ 
m_s(1\, GeV)= 165\, MeV\\ 
m_b(1\, GeV)= 5.4\, GeV
\end{array}\right]_{\mbox{\rm Ansatz-I}},\quad\quad\quad
\left[\begin{array}{lll}
m_u(1\, GeV)=1.0\, MeV   \\ 
m_c(1\, GeV)=1.4\, GeV \\
m_d(1\, GeV)=8.3\, MeV \\ 
m_s(1\, GeV)= 210\, MeV\\ 
m_b(1\, GeV)= 5.1\, GeV
\end{array}\right]_{\mbox{\rm Ansatz-II}}
$$
\bigskip

Output: $V_{CKM}=$

\begin{equation}
\left[
\begin{array}{lll}
0.9752 & 0.2213 & 0.0032 \\
0.2210 & 0.9745 & 0.0391 \\
0.0117 & 0.0375 & 0.9992 \\
\end{array}
\right]_{\mbox{\rm Ansatz-I}}
 ,\quad\quad\quad
\left[
\begin{array}{lll}
0.9754 & 0.2206 & 0.0030 \\
0.2205 & 0.9747 & 0.0377 \\
0.005 & 0.0374 & 0.9993 \\
\end{array}
\right]_{\mbox{\rm Ansatz-II}}
\label{result1}
\end{equation}

It is clear that in both ans\"atze, one obtains a good fit for the
experimentally observed $V_{CKM}$. The most salient difference between the 
two ans\"atze, is the fact that Ansatz-II requires a larger value for 
$m_s$ than Ansatz-I. This reflects the fact that Ansatz-I predicts in 
leading order $|V_{cb}|=\sqrt{2}\frac{m_s}{m_b}$, while Ansatz-II predicts
$|V_{cb}|=\frac{1}{\sqrt 2}\frac{m_s}{m_b}$ in leading order.

So far, we have only presented the predictions of our ans\"atze for the
moduli of $V_{CKM}$. Note that at present, with the exception of the CP
violating parameter $\epsilon$, all experimental results only measure or
put bounds on the moduli of $V_{CKM}$. In our ans\"atze one can readily
evaluate $J\equiv{\mbox{\rm Im}}(V_{12}V_{23}V_{13}^*V_{22}^*)$ which
measures the strength of CP violation. One obtains $J({\mbox{\rm
Ansatz-I}})=1.8\cdot 10^{-7}$, $J({\mbox{\rm Ansatz-II}})=0.8\cdot
10^{-6}$. These values are smaller than what is required to account for
the experimental value of $\epsilon$. Note however that in most extensions
of the SM there are new contributions to $\epsilon$. This is true for
example in the minimal supersymmetric standard model. 

\seccao{Renormalization Group Analysis}

In the previous section, we have tacitly assumed that our ans\"{a}tze are
implemented at 1 GeV. It is often advocated that one should look for a
fundamental theory of flavor at the unification scale. In this section, we
will analyse how our results for Ansatz-I change if we implement our
ans\"{a}tze at the unification scale. For that, we need to study the
renormalization group evolution of the CKM matrix and the quark masses.
The renormalization group equations (RGE) have been derived in a variety
of models. We will use the RGE for the Yukawa couplings in the case of the
SM with two Higgs doublets. More precisely, we will use the approximate
equations \cite{r7} for the diagonalized quark Yukawa coupling ratios
$Q^i_q$: 

\begin{equation}
16\pi^2\, \frac{d\, \log\, Q^i_q}{d\tau}\:=\: a_q \lambda_t^2+ 
 b_q \lambda_b^2 \qquad ; \quad q=t,b\quad ;\quad i=1,2
\label{eq51}
\end{equation}

\noi where $\tau=\log(\mu/M)$, $Q^1_t=\lambda_u/\lambda_t$,
$Q^2_t=\lambda_c/\lambda_t$, $a_t=b_b=3/2$, $a_b=b_t=1/2$ and the
$\lambda_j$ are the diagonalized Yukawa couplings. 

The evolution of $\lambda_t$ and $\lambda_b$ has to be calculated from the
original RGE of the Yukawa coupling matrices. However these reduce to

\begin{equation}
16\pi^2\, \frac{d\, \log\, \lambda_t}{d\tau}\:=C\,\lambda_t^2-
8\,g_3^2-\frac94\,g_2^2-\frac{17}{12}\,g_1^2
\label{eq52}
\end{equation}

\noi in the approximation where only one Yukawa coupling is dominant
\cite{r8}.  We will consider two possibilities. The first is assuming
strict universality, thus having in the Lagrangean prior to symmetry
breaking, only one universal coupling constant $\lambda$ as indicated in
Eq.(\ref{eq21}). This leads to $\lambda_b =\lambda_t$ and due to the fact
that $m_t\gg m_b$ this of course requires $v_u\gg v_d$. The other
possibility is assuming that the overall strength of Yukawa couplings is
different in the up and down quark sectors, leading to $\lambda_b\ll
\lambda_t$. These two possibilities lead to different values for C in
Eq.(\ref{eq52})

\begin{equation} 
C\:=\:5\quad ( \lambda_b =\lambda_t)
\qquad ;\qquad 
C\:=\:\frac92\quad ( \lambda_b\ll \lambda_t)
\label{eq53}
\end{equation}

In the study of the RGE for $V_{CKM}$, we will use a Wolfenstein-like
parametrization \cite{r9} defined in the following way: 

\begin{equation}
\begin{array}{c}
V_{12}=\lambda\, ; \quad
V_{23}=A\, \lambda^2\, ; \quad
V_{13}=A\mu \lambda^3 \exp(i\phi) \\ \\
V_{11},\: V_{22},\: V_{33}, \quad \quad \mbox{\rm real positive}
 \end{array}
\label{eq54}
\end{equation}

The advantage of this parametrization is that all parameters are simply and
exactly
related to measurable quantities, since the following definitions hold: 
$\lambda=|V_{12}|$, 
$A=|V_{23}/V_{12}^2|$, $\mu=|V_{13}/(V_{12}V_{23})|$ and 
$\phi=$arg$(V_{13} V_{22} V_{12}^\ast V_{23}^\ast)$. The only 
parameter with relevant evolution is $A$ whose RGE is given by:

\begin{equation}
16\pi^2\, \frac{d\, \log(A)}{d\tau}\:=-a\:\lambda^2_t
\label{eq55}
\end{equation}

\noi where $a=1$ when $\lambda_b =\lambda_t$ and $a=1/2$ when 
$\lambda_b\ll \lambda_t$.

We also need the RGE for $SU(3)\bigotimes SU(2)
\bigotimes U(1)$ gauge coupling constants for the case of the SM with two
Higgs doublets,

\begin{equation}
\frac{d\, \alpha_i}{d\tau}\:=-b_i\:\alpha_i^2\, ;
\quad \alpha_i=\frac{g_i^2}{4\pi}\quad\quad ;\quad i=3,2,1
\label{eq56}
\end{equation}

\noi where $b_{03}=11-2n_f/3$, $b_{02}=7-2n_f/3$, $b_{01}=-\frac13-10n_f/9$ 
and $n_f$ denotes number of flavors.

In order to make the integration of all RGE, Eqs.(\ref{eq51}),
(\ref{eq52}) and (\ref{eq55}), we have to start at some energy. As we
implement our ansatz at $M_X=10^{16}\, GeV$, the most logical assumption
would be to choose $M=M_X$ in $\tau=\log(\mu/M)$. In order to achieve this
we have to know the quark masses and gauge coupling constants at this
energy. For the light quarks $u$, $d$, $s$, $c$, $b$ we use the known
values at $1\,GeV$ and at a first step use the QCD running mass equations
to calculate the masses at $180\, GeV$, and then the Eq.(\ref{eq51}) to
obtain the quark ratios at $M_X$. For the gauge coupling constants we use
the values at $180\, GeV$ and Eq.(\ref{eq56}) to obtain their values at
$M_X$. 

Knowing the quark mass ratios at $M_X$, we implement our ansatz at this
scale. This means that we calculate the corresponding $V_{CKM}$ using
Eqs.(\ref{eq36a}) and (\ref{eq39}) for the phases $(r,q)$ and the
diagonalization formulas. We use then Eq.(\ref{eq55}) and the
parametrization described to evaluate $V_{CKM}$ down to $1\, GeV$.

\bigskip
Input:
\begin{equation}
\begin{array}{lllll}
m_u(1\, GeV)=1.0\, MeV & & m_d(1\, GeV)=6.5\, MeV & & 
\alpha_1(180\, GeV)=0.010 \\ 
m_c(1\, GeV)=1.35\, GeV & & m_s(1\, GeV)=165\, MeV & & 
\alpha_2(180\, GeV)=0.033 \\ 
& & m_b(1\, GeV)=5.4\, GeV & & \alpha_3(180\, GeV)=0.107 
\end{array}
\end{equation}
The physical top mass is chosen to be $174 \, GeV$.

Output:$V_{CKM}=$

\begin{equation}
\left(
\begin{array}{ccc}
0.9752 & 0.2212 & 0.0030 \\
0.2209 & 0.9745 & 0.0378 \\
0.0113 & 0.0362 & 0.9993 
\end{array}
\right)_{\lambda_b =\lambda_t}
\qquad
\left(
\begin{array}{ccc}
0.9752 & 0.2212 & 0.0033 \\
0.2209 & 0.9745 & 0.0395 \\
0.0120 & 0.0378 & 0.9992
\end{array}
\right)_{\lambda_b\ll\lambda_t}
\label{result2}
\end{equation}

Comparing the results of Eq.(\ref{result1}) with those of
Eq.(\ref{result2}), it is clear that the same input of quark masses at
$1\, GeV$, imposing our ansatz at $1\, GeV$ or at $M_X$, leads to similar
predictions for $V_{CKM}$. The only appreciable difference is in $V_{cb}$,
when we take $\lambda_b =\lambda_t$. 

\seccao{Comparison with texture zero approach}

In this section, we will address the question of whether there are some
common features between the USY ans\"{a}tze which we have constructed and
some of the texture structures classified in Ref.\cite{r2}. More
precisely, one may ask whether some of the USY ans\"{a}tze predict texture
zeros . We will not do an exhaustive study of the above question which is
rendered specially difficult due to the enormous freedom one has of making
weak-basis transformations which change the structure of Yukawa couplings
but do not alter their physical content. 

For definiteness, let us consider the following USY ansatz
\footnote{This is a slight variant of the ansatz of Eq.(\ref{eq312}), which 
leads to the same predictions for $V_{CKM}$ and it is more convenient for the 
analysis that follows}: 

\begin{equation}
M_{u,d}=c_{u,d}\left(  
\begin{array}{lll}
1 & e^{ir} & 1 \\
e^{iq} & 1 & e^{i(q+r)} \\
1 & 1 & 1
\end{array}
\right)_{u,d}
\label{eq61}   
\end{equation}  

We make now the following weak-basis transformations:

\begin{equation}
\begin{array}{r}
M_d \rightarrow M'_u=F^{\dagger}\cdot M_d \cdot K_u\cdot F \\ \\
M_d  \rightarrow M'_d=F^{\dagger}\cdot M_d  \cdot K_d\cdot F
\end{array}
\label{eq62}
\end{equation}

\noi where $K_u=\diag(1,e^{iq_u},1)$, $K_d=\mbox{\rm
diag}(1,e^{iq_d},1)$ and F is given by Eq.(\ref{eq314}). In the new basis
the mass matrices have the form: 

\begin{equation}
M'_{u,d}=3c_{u,d}\left(
\begin{array}{ccc}
0 & A & \frac A{\sqrt2} \\
-A & B & C \\
\frac {-A}{\sqrt2} & C & D
\end{array}
\right)_{u,d}  
\end{equation}

\noi where 

\begin{equation}
\begin{array}{l}
A=\frac1{3\sqrt3}e^{iq}[e^{ir}-1]\cong \frac i{3\sqrt3} (r) 
\\ \\
B=\frac29[1-e^{i(q+r)}]\cong\frac{-2i}9 (q+r) 
\\ \\
C=\frac{-4+3e^{iq}+e^{i(q+r)}}{9\sqrt2} \cong
\frac i{9\sqrt2}[2q+r]
\\ \\
D=\frac{4+3e^{iq}+2e^{i(q+r)}}9\cong 1
\end{array}
\end{equation}

Note that $M'_u$, $M'_d$ have the same structure and both exhibit zeros in
the (1,1) element. It is interesting to note that all the texture zero
structures classified in Ref.\cite{r2} also have zeros in the (1,1)
position. Note that $M'_{u,d}$ are written in the so called heavy-basis,
in the sense that after factoring out the over-all constant $3c_{u,d}$,
the moduli of all the elements of $M'_{u,d}$ are much smaller than 1,
except the element (3,3). At this stage one may ask whether it is possible
to make further weak-basis transformations, leading to other texture
zeros. Since for $M'_{u,d}$ the following relations hold: 

\begin{equation}
(M'_{u,d})_{12}=\sqrt2\,\,(M'_{u,d})_{13}
\qquad \qquad
(M'_{u,d})_{31}=\sqrt2\,\,(M'_{u,d})_{21}
\label{eq64}
\end{equation}

\noi one may be tempted to make the weak-basis transformation
\footnote{This weak-basis transformation was pointed out to us by Daniel 
Felizardo and Jo\~ao Seixas}:

\begin{equation}
M'_{u,d}\rightarrow M''_{u,d}=O^T\cdot M'_{u,d}\cdot O
\end{equation}

\noi where

\begin{equation}
O=\left(
\begin{array}{ccc}
1 & 0 & 0 \\
0 & \cos\:\theta & -\sin\:\theta \\
0 & \sin\:\theta & \cos\:\theta
\end{array}
\right)
\qquad \qquad \theta=\arctan(\frac1{\sqrt2})
\end{equation}

The above weak-basis transformation will indeed lead to zeros in the
elements (1,3), (3,1) of $M''_{u,d}$ while maintaining the zero in (1,1). 
However the matrices $M''_{u,d}$ are no longer written in a heavy-basis
and thus a comparison with the texture zero structures of Ref.\cite{r2}
looses its meaning. 

To conclude, in the case of the USY ansatz of Eq.(\ref{eq61}), one has in
the heavy-basis, a texture zero in the element (1,1) in both $M'_u$ and
$M'_d$. This USY ansatz is further characterized by a simple relation 
among some of the matrix elements, given by Eq.(\ref{eq64})

\seccao{Conclusions}

The idea that the flavor structure of Yukawa couplings is all contained 
in their phases is intriguing. The limit $m_u=m_d=0$ is specially 
interesting in USY, since all solutions correspond to simple choices for 
the USY phases and can be readily classified. We have explored this fact 
to make a systematic search for calculability of the CKM matrix elements 
in terms of quark ratios. It was pointed out that within USY a natural 
explanation is found for the mixing between the first two generations being 
significantly larger than other CKM mixings. 
The ans\"atze we presented have a highly predictive power, since Ansatz-I 
predicts $V_{CKM}$ in terms of quark mass ratios with no free parameters 
and Ansatz-II predicts $V_{CKM}$ with only one free parameter.
The fact that the 
experimentally observed CKM matrix can be accommodated within USY 
ans\"atze corresponding to simple relations among the phases, makes the 
USY hypothesis specially appealing.

\pagebreak

\appendix
\noi   {\Large \bf Appendix}
\def\theequation{A\arabic{equation}}

\bigskip

\noi In this appendix we show some rather peculiar features of USY. 
In part (a), we consider the USY hypothesis for the case of two generations 
and analyse the question of calculability of the Cabibbo angle in terms 
quark mass ratios. Following the approach used for three generations, 
we will show that for two generations USY leads either to an arbitrary 
non-calculable  Cabibbo angle  or to the unrealistic relation 
$|\theta_C|=\frac {m_d}{m_s}\pm\frac {m_u}{m_c}$. 

\noi In part (b), we will show that the experimentally observed quark masses 
together with the USY hypothesis, necessarily imply $V_{CKM}\ne\UM$, 
independently of all USY parameters.
\bigskip 

\noi a) {\bf USY in two generations}

For the case of two generations, an appropriate weak basis transformation 
with diagonal unitary phase matrices, analogues of Eq.(\ref{eq22}), transforms 
the up and down quark mass matrices without loss of generality, into

\begin{equation}
M_u=c_u\: \left(
\begin{array}{cc}
e^{ip_u} & 1 \\
1  & 1
\end{array}
\right) ,
\quad
M_d\:=\:c_d \:
\left(
\begin{array}{cc}
1 & 0 \\
0  & e^{i\alpha}
\end{array}
\right)
\cdot
\left(
\begin{array}{cc}
e^{ip_d} & 1 \\
1  & 1
\end{array}
\right)
\cdot
\left(
\begin{array}{cc}
1 & 0 \\
0  & e^{-i\alpha}
\end{array}
\right)
\label{eq91}
\end{equation}

It is clear that the phase $\alpha$ does not affect the quark spectrum. 
In an attempt to reach calculability of the Cabibbo angle and following 
our approach for three generations, we set the phase $\alpha$ to
zero. The $M_{u,d}$ mass matrix is then of the 
form

\begin{equation}
M=c\:\left(
\begin{array}{cc}
e^{ip} & 1 \\
1  & 1
\end{array}
\right)
\label{eq92}
\end{equation}

We have the following mass spectrum

\begin{equation}
\begin{array}{llll}
m^2_u=4c_u^2\,\sin^2(\frac{p_u}4) & &
m^2_d=4c_d^2\,\sin^2(\frac{p_d}4)  \\
m^2_c=4c_u^2\,\cos^2(\frac{p_u}4) & &
m^2_s=4c_d^2\,\cos^2(\frac{p_d}4) 
\end{array}
\label{eq93}
\end{equation}

These relations fix the phases $p_u$ and $p_d$ in terms of the quark mass 
ratios,

\begin{equation}
\tan^2(\frac{p_u}4)\:=\:\frac{m_u^2}{m_c^2} 
\qquad ,\qquad
\tan^2(\frac{p_d}4)\:=\:\frac{m_d^2}{m_s^2}
\label{eq94}
\end{equation}

The Cabibbo angle is obtained by diagonalizing the mass matrix 
$H=MM^\dagger$, through the unitary matrix  

\begin{equation}
U=\left(
\begin{array}{cc}
1 & 0 \\
0 & e^{-i\frac{p}{2}}
\end{array}
\right)\cdot
\left(
\begin{array}{cc}
\frac1{\sqrt2} & \frac1{\sqrt2} \\
\frac{-1}{\sqrt2} & \frac1{\sqrt2}
\end{array}
\right)
\label{eq95}
\end{equation}

\noi The $V_{CKM}$ matrix is given by:

\begin{equation}
V_{CKM}=\left(
\begin{array}{cc}
\cos\,\theta_C & -\sin\,\theta_C \\
\sin\,\theta_C & \cos\,\theta_C
\end{array} \right)
\label{eq96}
\end{equation}

\noi where we have eliminated unphysical phases in $V_{CKM}$. 
The Cabibbo angle $\theta_C$ iscara given in leading order by

\begin{equation}
|\theta_C|\:=\:|\frac{p_d}4-\frac{p_u}4|
\label{eq97}
\end{equation}

Combining this equation with Eq.(\ref{eq94}), we find in leading order

\begin{equation}
|\theta_C|\:=\:\frac{m_d}{m_s}\pm \frac{m_u}{m_c}
\label{eq98}
\end{equation}

\noi the signs depend on the square roots which have to be calculated from 
Eq.(\ref{eq94}).

In connection with this result we make the following observations:

\noi i) If the diagonal matrix $K=\diag(1,e^{i\alpha})$ in Eq.(\ref{eq91})
is present with $\alpha$ arbitrary, then the relation of Eq.(\ref{eq98})
is lost. In this case there is no correlation between the masses and the
Cabibbo angle\footnote{This was mentioned in Ref.[3]}. 

\noi ii) The fact that for two generations one obtains the 
$|\theta_C|\approx m_d/m_s$ instead of $|\theta_C|\approx \sqrt{m_d/m_s}$ 
can be viewed in the context of the USY hypothesis as an indication 
that there are more than two generations. 
We find quite intriguing the 
fact that in USY, the successful relation $|V_{us}|=\sqrt{\frac {m_d}{m_s}}$ 
naturally arises for three generations.
\bigskip\bigskip

\noi b) {\bf  Quark spectrum and $V_{CKM}\ne\UM$ in USY }

\smallskip We will show that in three generations USY one necessarily has
$V_{CKM}\ne\UM$, since $V_{CKM}=\UM$ would imply an unrealistic quark mass
spectrum. 

\noi Note that $V_{CKM}=\UM$ implies

\begin{equation}
\left[H_u\, , \, H_d \right]\:=\:0
\label{eq99}
\end{equation}

\noi where $H_u$ and $H_d$ are defined as in Eq.(\ref{eq23a}). 
Writing for all practical purposes $H_{u,d}$ in the form

\begin{equation}
H_u\:=\: \left(
\begin{array}{ccc}
1 & x_1e^{-i\,\alpha_1} &  x_2e^{-i\,\alpha_2} \\
x_1e^{i\,\alpha_1} & 1 & x_3e^{-i\,\alpha_3} \\
x_2e^{i\,\alpha_2} & x_3e^{i\,\alpha_3} & 1
\end{array}
\right)
\label{eq910}
\end{equation}

\noi one concludes from Eq.(\ref{eq99}), that

\begin{equation}
\frac{x_i^u}{x_i^d}\:=\: \frac{x_j^u}{x_j^d} \qquad 
\alpha^u_i\:=\:\alpha^d_i \quad \quad i,j=1,2,3
\label{eq911}
\end{equation}

By inserting these relations into the characteristic equations of $H$ for 
the determinant, $\delta$, and second invariant, $\chi$,

\begin{equation}
\begin{array}{l}
det(H)=\delta=1+2 x_1\, x_2\, x_3\, \cos(\alpha_1+\alpha_3-\alpha_2)%
-x_1^2-x_2^2-x_3^2 \\ \\
\chi(H)=\chi=3-x_1^2-x_2^2-x_3^2
\end{array}
\label{eq912}
\end{equation}

\noi one finds

\begin{equation}
\frac{x_1^u\, x_2^u\, x_3^u}{x_1^d\, x_2^d\, x_3^d}\:=\:
\frac{1-\frac{\chi_u-\delta_u}2}{1-\frac{\chi_d-\delta_d}2}
\label{eq913}
\end{equation}

\begin{equation}
\left( \frac{x_1^u}{x_1^d}\right)^2\,
\left[ 
\frac{1+\left(\frac{x_2^u}{x_1^u}\right)^2+\left(\frac{x_3^u}{x_1^u}\right)^2}%
{1+\left(\frac{x_2^d}{x_1^d}\right)^2+\left(\frac{x_3^d}{x_1^d}\right)^2}
\right] \:=\: \frac{1-\frac{\chi_u}3}{1-\frac{\chi_d}3}
\label{eq914}
\end{equation}

Finally, combining Eqs.(\ref{eq911}), (\ref{eq913}) and (\ref{eq914}), 
one gets an exact relation between the masses 
of the up and down sector,

\begin{equation}
\frac{\left( 1-\frac{\chi_u}3 \right)^{\frac32}}{1-\frac{\chi_u-\delta_u}2}
\:=\:
\frac{\left( 1-\frac{\chi_d}3 \right)^{\frac32}}{1-\frac{\chi_d-\delta_d}2}
\label{eq915}
\end{equation}

\noi Calculating the Taylor-MacLaurin series on both sides of this 
equation, yields in first order, $\chi_u^2=\chi_d^2$, thus implying that,
$m_t\:=\:m_c\:\frac{m_b}{m_s}$,
which is in clear disagreement with the experimental value of the top quark
mass.

\pagebreak

{\bf \large Table 1}

\begin{displaymath}
\begin{array}{|l|ll|l|l|}
\hline
1) & p=0 & s=q+r & \va12\raiz{m_1}{m_2} &
 \va13\frac1{\sqrt2}\raiz{m_1 m_2}{m_3^2} \\ \cline{4-5}
& \dd{r} & \cc{q+r} & \va31\frac3{\sqrt2}\raiz{m_1 m_2}{m_3^2} & 
\va23\sqrt2\frac{m_2}{m_3} \\ \cline{4-5}
\hline \hline 
%%%%%%%%%%%%%%%%%%%%%%%%%%%%%%%%
2) & s=0 & r=-q & \va12\raiz{m_1}{m_2} & 
\va13\frac1{\sqrt2}\raiz{m_1 m_2}{m_3^2} \\ \cline{4-5}
& \dd{q} & \cc{p} & \va31\sqrt\frac32\frac{m_1}{m_3} & 
\va23\frac1{\sqrt2}\frac{m_2}{m_3}(1-\sqrt3\raiz{m_1}{m_2}) \\ \cline{4-5}
\hline \hline 
%%%%%%%%%%%%%%%%%%%%%%%%%%%%%%%%
3) & s=q & q=-p & \va12\raiz{m_1}{m_2} & 
\va13\sqrt2\raiz{m_1 m_2}{m_3^2} \\ \cline{4-5}
& \dd{p} & \cc{r-p} & \va31\frac3{\sqrt2}\raiz{m_1 m_2}{m_3^2} & 
\va23\frac1{\sqrt2}\frac{m_2}{m_3}(1+2\sqrt3\raiz{m_1}{m_2}) \\ \cline{4-5}
\hline \hline 
%%%%%%%%%%%%%%%%%%%%%%%%%%%%%%%%
4a) & s=-r & r=\frac{p-q}2 & \va12\raiz{m_1}{m_2} & 
\va132\sqrt2\raiz{m_1 m_2}{m_3^2} \\ \cline{4-5}
& \dd{\frac{p-q}2} & \cc{\frac{p+q}2} & 
\va31\frac3{\sqrt2}\raiz{m_1 m_2}{m_3^2} & 
\va23\frac1{\sqrt2}\frac{m_2}{m_3} \\ \cline{4-5}
\hline \hline 
%%%%%%%%%%%%%%%%%%%%%%%%%%%%%%%
4b)& s=-r & q=-p & \va12\raiz{m_1}{m_2} & 
\va13\sqrt2\raiz{m_1 m_2}{m_3^2} \\ \cline{4-5}
& \dd{r} & \cc{p} & \va31\frac3{\sqrt2}\raiz{m_1 m_2}{m_3^2} 
& \va23\frac1{\sqrt2}\frac{m_2}{m_3} \\ \cline{4-5}
\hline
\end{array}
\end{displaymath}

\bigskip
{\bf Acknowledgement}

We would like to thank M.N. Rebelo for a careful reading of the manuscript.

\end{document}